\documentclass{article}
\usepackage{amsmath}
\usepackage{epsfig}
\usepackage{amssymb}
\usepackage{fullpage}

\def\##1{{\underline #1}}
\def\~#1{{\underline {\mathcal #1}}}
\def\+#1{{{\mathcal #1}}}
\def\=#1{\underline{\underline #1}}

\def\.{\mbox{ \tiny{$^\bullet$} }}

\def\eps{\epsilon}
\def\epso{\epsilon_o}

\def\muo{\mu_o}
\def\ko{k_o}

\def\ux{\#u_x}
\def\uy{\#u_y}
\def\uz{\#u_z}
\def\muo{\mu_o}
\def\ko{k_o}

\def\vkap{\varkappa}

\begin{document}

\begin{center}{\large\bf ELECTRICAL CONTROL OF SURFACE--WAVE PROPAGATION AT THE PLANAR INTERFACE OF A LINEAR ELECTRO--OPTIC MATERIAL AND AN ISOTROPIC
DIELECTRIC MATERIAL 
 }\end{center} \vskip 1 cm

\noindent SUDARSHAN R. NELATURY\\
Department of Electrical, Computer and Software Engineering, \\
Pennsylvania State University, The Behrend College,\\
5101 Jordan Road, Erie, PA 16563-1701, USA.\\

\noindent JOHN A. POLO JR.\\
Department of Physics and Technology,\\
Edinboro University of Pennyslvania,\\
235 Scotland Rd., Edinboro, PA  16444, USA.\\

\noindent AKHLESH LAKHTAKIA\\
CATMAS~---~Computational \& Theoretical Materials Sciences Group, \\
Department of Engineering Science and
Mechanics, \\
Pennsylvania State University, University Park, PA  16802, USA.

\begin{abstract}
Surface waves can propagate on the planar interface of a linear electro-optic (EO) material and an
isotropic dielectric material, for restricted ranges of the orientation angles of the EO material and
the refractive index of the isotropic material. These ranges can be controlled by the application of a
dc electric field, and depend on both the magnitude and the direction of the dc field. Thus,
surface-wave propagation can be electrically controlled by exploiting the Pockels effect.

\end{abstract}

\section{Introduction}
Wave propagation localized to a planar metal-dielectric interface has a long history
(Zenneck, 1907; Agranovich and Mills, 1982; Boardman, 1982; Homola
et al., 1999; Matveeva et al., 2005).
 In contrast,
wave propagation localized to the
planar interface of an isotropic dielectric material and a uniaxial dielectric material
was found possible by D'yakonov only in  1988 (D'yakonov, 1988; Averkiev and
Dyakonov, 1990). Since then,  researchers have explored surface-wave propagation (SWP) on
increasingly complex systems of bimaterial
interfaces such as biaxial-isotropic, uniaxial-uniaxial, and biaxial-biaxial
(Walker et al., 1998; Darinski\u{i} 2001, Wong et al.,
2005; Polo et al., 2006, 2007a,b; Nelatury et al., 2007). In all such investigations,
essentially, one looks for the selected angular regimes of the propagation direction wherein certain dispersion
conditions are met. For complex systems, the angular regimes are very
narrow and depend highly on the crystallographic symmetries of the two materials.

Our motivation for this paper is to show how one might exercise control on the angular regimes of SWP. Whereas temperature and pressure may be altered to
control SWP, the application of an external field is expected to provide dynamic
control.
In particular, the
electro-optic  effect refers to changes in optical properties by the application of a low-frequency or dc
electric field. For instance, an optically isotropic crystal (possessing either
the $\bar{4}3m$ or $23$ point group symmetry) upon exposure to a dc electric
field turns  birefringent (Cook, 1996; Lakhtakia, 2006a).

The Pockels effect is a linear electro-optic (EO) effect, whereby the modification of the inverse (optical)
permittivity matrix by the dc electric field is quantified through 18 electro-optic coefficients, not
all of which may be independent of each other, depending on the point group symmetry (Boyd, 1992). The
opportunities offered by the Pockels effect for tuning the optical response characteristics of materials
have recently been highlighted for photonic band-gap engineering (Lakhtakia, 2006b; Li et al., 2007)
and composite materials (Lakhtakia and Mackay, 2007; Mackay and Lakhtakia, 2007). These publications
suggest that, although the changes in the optical permittivity matrix are typically small,
 the effect on SWP could be significant due to the extreme
sensitivity of surface waves on the dielectric constitutive properties. Let us note here an earlier
study wherein the quadratic EO effect named after Kerr was shown to offer control over SWP (Torner et
al., 1993).

This paper introduces the influence of the Pockels effect on SWP. Although
the theoretical treatment is general, numerical results are presented only for
a specific EO material:  potassium niobate (Zgonik et al., 1993). The remainder of the paper is organized as follows: Section 2 provides a
description of a canonical boundary-value problem, the optical permittivity matrix of an EO crystal, and the
derivation of the dispersion relations for SWP. In Section 3 numerical results are furnished, and our
conclusions are distilled in Section 4. A note on notation: vectors are underlined, matrixes are
decorated with an overbar,  and the
Cartesian unit vectors are denoted as $\ux$, $\uy$, and $\uz$. All field quantities are assumed to have
an $\exp({-i\omega t})$ time-dependence.

\section{Theory}
Let the plane of SWP be   a bimaterial interface. The half-space $z< 0$ is filled with a homogeneous,
isotropic, dielectric material with an optical refractive index denoted by $n_s$. The half-space $z> 0$
is filled with a homogeneous, linear, EO material, whose optical relative permittivity matrix  is stated
as (Lakhtakia and Reyes 2006a,b)
\begin{equation}
\bar{\epsilon}_{\,rel} = \bar{S}_{z}\left(\psi\right)\cdot
\bar{R}_{y}(\chi) \cdot\bar{\epsilon}_{PE} \cdot\bar{R}_{y}(\chi)\cdot
\bar{S}_{z}\left(-\psi\right)\,.  \label{AAepsr}
\end{equation}
Incorporating the Pockels effect due to an arbitrarily oriented  but uniform dc electric field
${\#E}^{dc} $, the matrix $\bar{\epsilon}_{PE}$  is given by
\begin{equation}
\displaystyle{\bar{\epsilon}_{PE}\approx\left(
\begin{array}{ccc}
\epsilon _{1}^{(0)}(1-\epsilon _{1}^{(0)}\sum_{K=1}^3 r_{1K}E_{K}^{dc} ) &
-\epsilon _{1}^{(0)}\epsilon _{2}^{(0)}\sum_{K=1}^3 r_{6K}E_{K}^{dc} &
-\epsilon _{1}^{(0)}\epsilon _{3}^{(0)}\sum_{K=1}^3 r_{5K}E_{K}^{dc} \\[5pt]
-\epsilon _{2}^{(0)}\epsilon _{1}^{(0)}\sum_{K=1}^3 r_{6K}E_{K}^{dc} &
\epsilon _{2}^{(0)}(1-\epsilon _{2}^{(0)}\sum_{K=1}^3 r_{2K}E_{K}^{dc} ) &
-\epsilon _{2}^{(0)}\epsilon _{3}^{(0)}\sum_{K=1}^3 r_{4K}E_{K}^{dc} \\[5pt]
-\epsilon _{3}^{(0)}\epsilon _{1}^{(0)}\sum_{K=1}^3 r_{5K}E_{K}^{dc} &
-\epsilon _{3}^{(0)}\epsilon _{2}^{(0)}\sum_{K=1}^3 r_{4K}E_{K}^{dc} &
\epsilon _{3}^{(0)}(1-\epsilon _{3}^{(0)}\sum_{K=1}^3 r_{3K}E_{K}^{dc} )
\end{array}
\right) }\,,  \label{PocEps}
\end{equation}
correct to the first order in $\vert\#E^{dc}\vert$,
where
\begin{equation}
\left(
\begin{array}{l}
E_{1}^{dc} \\[5pt]
E_{2}^{dc} \\[5pt]
E_{3}^{dc}
\end{array}
\right) = \bar{R}_{y}(\chi)\cdot\bar{S}_{z}\left(-\psi\right)\cdot
\left(
\begin{array}{l}
E_{x}^{dc} \\[5pt]
E_{y}^{dc} \\[5pt]
E_{z}^{dc}
\end{array}
\right) \,,
\end{equation}
$\epsilon _{1,2,3}^{(0)}$ are the principal relative permittivity scalars in
the optical regime, whereas $r_{JK}$ (with $1\leq J\leq 6$ and $1\leq K\leq
3 $) are the EO coefficients.
The EO material can be  isotropic, uniaxial, or biaxial, depending on the
relative values of $\epsilon_1^{(0)}$, $\epsilon_2^{(0)}$, and $%
\epsilon_3^{(0)}$. Furthermore, the EO material may belong to one of 20 crystallographic classes of
point group symmetry, in accordance with the relative values of the EO coefficients $r_{JK}$.

\begin{figure}
    \begin{center}
    \begin{tabular}{c}
    \includegraphics[height=7cm]{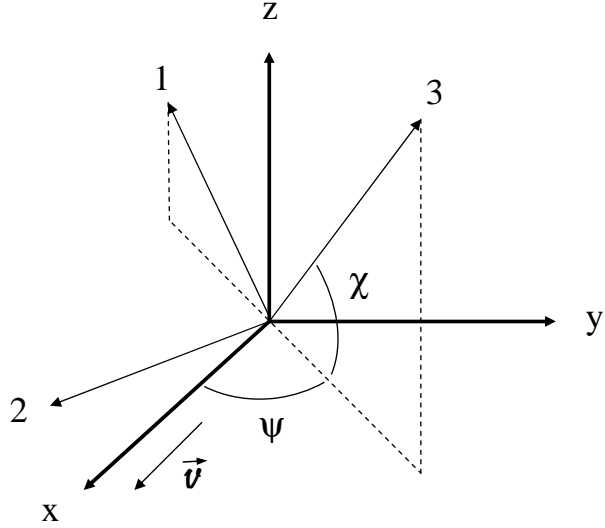}
    \end{tabular}
    \end{center}
    \caption{ \label{Fig:Geometry} Relationship of laboratory coordinates $(x,y,z)$ to
    the conventional electro-optic coordinates $(1,2,3)$. SWP is taken
    to occur parallel to the $x$ axis with phase velocity $\#v$.}
 \end{figure}

The rotation matrix
\begin{equation}
\bar{S}_z(\psi)=\left(
\begin{array}{ccc}
\cos \psi & -\,\sin\psi & 0 \\
\sin\psi & \cos \psi & 0 \\
0 & 0 & 1
\end{array}
\right)
\end{equation}
in Equation (\ref{AAepsr}) denotes a rotation about the $z$ axis by an angle $\psi \in\left[0,2\pi\right)$. The matrix
\begin{equation}
\bar{R}_{y}(\chi )=\left(
\begin{array}{ccc}
-\sin \chi & 0 & \cos \chi \\
0 & -1 & 0 \\
\cos \chi & 0 & \sin \chi
\end{array}
\right)
\end{equation}
involves the angle $\chi \in\left[0,\pi/2\right]$ with respect to the $x$ axis in the $xz$
plane, and combines a rotation as well as an inversion.  The angles $\psi$ and $\chi$ delineate the orientation
of the EO material in the laboratory coordinate system, the full
transformation from laboratory coordinates ($x,y,z$) to those used conventionally for EO materials
($1,2,3$) being illustrated in Figure~\ref{Fig:Geometry}.

Now $\bar{\epsilon}_{PE}$ is a symmetric matrix, regardless of the magnitude and direction of
$\#E^{dc}$. With the assumption that all of its elements are real-valued,  $\bar{\epsilon}_{PE}$ can be
written as $\alpha_1\, (\ux\ux+\uy\uy+\uz\uz) +\alpha_2\, (\#u_m\#u_n+\#u_n\#u_m)$, where  $\alpha_1$ and $\alpha_2$ are
scalars and  the  unit vectors $\#u_m$ and $\#u_n$ are parallel to the crystallographic axes or the
optical ray axes of the EO material in the ($1,2,3$) coordinate system (Chen 1983). Likewise,
$\bar{\epsilon}_{PE}^{-1}$ can be written as $\beta_1 \,(\ux\ux+\uy\uy+\uz\uz) +\beta_2\, (\#u_p\#u_q+\#u_q\#u_p)$, where
$\beta_1$ and $\beta_2$ are scalars and  the  unit vectors $\#u_p$ and $\#u_q$ are parallel to the optic
axes of the EO material in the ($1,2,3$) coordinate system. Thus, the application of the uniform dc
electric field not only changes the eigenvalues of $\bar{\epsilon}_{PE}$, but also rotates both
optic axes and both optical ray axes,
in general.

\subsection{Field representations}

Without loss of generality we assume that the SWP direction   is parallel to the $x$ axis. The fields in
the half-space $z<0$ must satisfy the equations
\begin{equation}\left.\begin{array}{l}
\#k_s \times \~E_s=\omega\muo\,\~H_s\\[4pt]
\#k_s \times \~H_s=-\omega\eps_o\, n_s^2\,
\~E_s\end{array}\right\}\,, \label{EH-s}
\end{equation}
where $\epso$ and $\muo$ are the permittivity and permeability of free space, and $\~E_s$ and $\~H_s$
are the complex-valued amplitudes of the electric and magnetic field phasors in the isotropic dielectric material. The wave
vector
\begin{equation}
\#k_s=\ko\,(\vkap\, \ux- iq_s\, \uz)\,, \label{kSub}
\end{equation}
where $\vkap$ and
\begin{equation}
q_s=+\sqrt{\vkap^2-n_s^2}\,\label{eqn:dispSub}
\end{equation}
are the normalized propagation constant and the decay constant, respectively, whereas
$\ko=\omega\sqrt{\epso\muo}$ is the free-space wavenumber. We must have ${\rm Re}\left[q_s\right]>0$ for
SWP; furthermore, $\vkap$ must be real-valued and positive for un-attenuated propagation along the $x$ axis. Accordingly,
the acceptable solutions of Equations~(\ref{EH-s}) are
\begin{equation}
\~E_s= A_{s1}\,\uy + A_{s2}\left( {iq_s}\,\ux +\vkap\,\uz\right)
\label{eqn:Esub}
\end{equation}
and
\begin{equation}
\~H_s=\sqrt{\frac{\epso}{\muo}}\left[ A_{s1}\left(i q_s\,\ux +\vkap\,\uz\right)-
A_{s2}\,{n_s^2}\uy\right]\,,\label{eqn:Hsub}
\end{equation}
where $A_{s1}$ and $A_{s2}$ are unknown scalars. If $q_s$ is purely real-valued, the electric and
magnetic field phasors of the surface wave decay exponentially with respect to $z$ as $z\to-\infty$;
otherwise, the decay is damped sinusoidal.

The fields in the half-space $z>0$ must be solutions of the equations
\begin{equation}\left.\begin{array}{l}
\#k_e \times \~E_e=\omega\muo\,\~H_e\\[4pt]
\#k_e \times \~H_e=-\omega\eps_o\,\bar{\epsilon}_{\,rel}\cdot\~E_e\end{array}\right\}\,. \label{eqn:curlEH}
\end{equation}
The wave vector
\begin{equation}
\#k_e=\ko\,( \vkap\, \ux+ iq_e\, \uz)\, \label{kPE}
\end{equation}
must have ${\rm Re}\left[q_e\right]>0$ for localization of energy to the bimaterial interface. A purely
real-valued $q_e$ indicates an exponential decay of the field phasors with respect to $z$ as
$z\to+\infty$, whereas ${\rm Im}\left[q_e\right] \ne 0$ indicates a damped-sinusoidal decay.

Substitution of  $\#k_e$  into Equations~(\ref{eqn:curlEH}) yields a set of six homogenous equations that are linear
in the six Cartesian components of $\~E_e$ and $\~H_e$.  Setting the determinant of the coefficients equal to zero
gives the dispersion equation
\begin{equation}
C_0+C_1q_e+C_2q_e^2+C_3q_e^3+C_4q_e^4=0\,,\label{eqn:dispEO}
\end{equation}
where the coefficients
\begin{eqnarray}
\nonumber
C_0&=&\eps_{xx}\eps_{yz}^2+\eps_{xy}^2\eps_{zz}+\eps_{xz}^2\eps_{yy}-\eps_{xx}\eps_{yy}\eps_{zz}
     -2\eps_{xy}\eps_{xz}\eps_{yz}\\[5pt]
&&+(\eps_{xx}\eps_{yy}+\eps_{xx}\eps_{zz}-\eps_{xy}^2-\eps_{xz}^2)\vkap^2 -\eps_{xx}\vkap^4\,,\\[5pt]
C_1&=&2i(\eps_{xz}\eps_{yy}-\eps_{xy}\eps_{yz})\vkap-2i\eps_{xz}\vkap^3\,,\\[5pt]
C_2&=&\eps_{xz}^2+\eps_{yz}^2-(\eps_{xx}+\eps_{yy})\eps_{zz}+(\eps_{xx}+\eps_{zz})\vkap^2\,,\\[5pt]
C_3&=&2i\vkap\eps_{xz}\,,\\[5pt]
C_4&=&-\eps_{zz}\,,
\end{eqnarray}
involve $\epsilon_{xy}=\ux\cdot\bar{\epsilon}_{rel}\cdot\uy$, etc.

The solution of Equation~(\ref{eqn:dispEO}) leads to four values of $q_e$. We select the two values of
$q_e$ that conform to the restriction ${\rm Re}\left[q_e\right]>0$, and label them as $q_{e1}$ and
$q_{e2}$. The corresponding wave vectors in the EO material are denoted by $\#k_{e1}$ and $\#k_{e2}$.
Accordingly, with unknown coefficients $A_{e1}$ and $A_{e2}$, we write the Cartesian components of the
field phasors in the half-space $z>0$ as
\begin{eqnarray}
\nonumber
 \ux\cdot\~E_e &=&\frac{\eps_{xz}\eps_{yy}-\eps_{xy}\eps_{yz}+\eps_{xz}q_{e1}^2+i(\eps_{yy}q_{e1}+q_{e1}^3)\vkap-i(\eps_{xz}+q_{e1})\vkap^2}
 {-\eps_{xy}\eps_{xz}+\eps_{xx}\eps_{yz}+\eps_{yz}q_{e1}^2-i\eps_{xy}q_{e1}\vkap}A_{e1}\\[5pt]
 &&+\frac{\eps_{xz}\eps_{yy}-\eps_{xy}\eps_{yz}+\eps_{xz}q_{e2}^2+i(\eps_{yy}q_{e2}+q_{e2}^3)\vkap-i(\eps_{xz}+q_{e2})\vkap^2}
 {-\eps_{xy}\eps_{xz}+\eps_{xx}\eps_{yz}+\eps_{yz}q_{e2}^2-i\eps_{xy}q_{e2}\vkap}A_{e2}\,,\\[5pt]
 \uy\cdot
\~E_e &=& A_{e1}+A_{e2}\,,\\[5pt]
\nonumber
\uz\cdot \~E_e
 &=&\frac{\left(\eps_{xx}\eps_{yy}-\eps_{xy}^2+(\eps_{xx}+\eps_{yy})q_{e1}^2+q_{e1}^4\right)-(\eps_{xx}+q_{e1}^2)\vkap^2}
 {-\eps_{xx}\eps_{yz}+\eps_{xy}\eps_{xz}-\eps_{yz}q_{e1}^2+i\eps_{xy}q_{e1}\vkap}A_{e1}\\[5pt]
&&\frac{\left(\eps_{xx}\eps_{yy}-\eps_{xy}^2+(\eps_{xx}+\eps_{yy})q_{e2}^2+q_{e2}^4\right)-(\eps_{xx}+q_{e2}^2)\vkap^2}
 {-\eps_{xx}\eps_{yz}+\eps_{xy}\eps_{xz}-\eps_{yz}q_{e2}^2+i\eps_{xy}q_{e2}\vkap}A_{e2}\,,\\[5pt]
 \nonumber
 \mathrm{and}&&\\[5pt]
\ux\cdot \~H_e &=&-i\sqrt{\frac{\epso}{\mu_o}} q_{e1}(A_{e1}+A_{e1})\,,\\[5pt]
\uy\cdot \~H_e  &=&i\sqrt{\frac{\epso}{\muo}}\left\{\frac{i(-\eps_{xy}\eps_{yz}+\eps_{xz}\eps_{yy})q_{e1}+\eps_{xz}q_{e1}^3+(-\eps_{xx}+\eps_{xx}\eps_{yy}-\eps_{xy}^2)\vkap}
 {\eps_{xx}\eps_{yz}-\eps_{xy}\eps_{xz}+\eps_{yz}q_{e1}^2-i\eps_{xy}q_{e1}\vkap}\right.\\[5pt]
 &&+\left.\frac{(-\eps_{xx}+\eps_{xx}\eps_{yy}-\eps_{xy}^2-i\eps_{xz}q_{e1}+\eps_{xx}q_{e1}^2)\vkap^3-\eps_{xx}\vkap^4}
 {\eps_{xx}\eps_{yz}-\eps_{xy}\eps_{xz}+\eps_{yz}q_{e1}^2-i\eps_{xy}q_{e1}\vkap}\right\}A_{e1}\\[5pt]
 &&+i\sqrt{\frac{\epso}{\muo}}\left\{\frac{i(-\eps_{xy}\eps_{yz}+\eps_{xz}\eps_{yy})q_{e2}+\eps_{xz}q_{e2}^3+(-\eps_{xx}+\eps_{xx}\eps_{yy}-\eps_{xy}^2)\vkap}
 {\eps_{xx}\eps_{yz}-\eps_{xy}\eps_{xz}+\eps_{yz}q_{e2}^2-i\eps_{xy}q_{e2}\vkap}\right.\\[5pt]
 &&+\left.\frac{(-\eps_{xx}+\eps_{xx}\eps_{yy}-\eps_{xy}^2-i\eps_{xz}q_{e2}+\eps_{xx}q_{e2}^2)\vkap^3-\eps_{xx}\vkap^4}
 {\eps_{xx}\eps_{yz}-\eps_{xy}\eps_{xz}+\eps_{yz}q_{e2}^2-i\eps_{xy}q_{e2}\vkap}\right\}A_{e2}\,,\\[5pt]
 \uz\cdot\~H_e  &=&\sqrt{\frac{\epso}{\muo}}\vkap (A_{e1}+ A_{e2})\,.
\end{eqnarray}

\subsection{Boundary conditions}
The boundary conditions at the interface $z=0$ lead to the following four
equations:
\begin{equation}
\left.
\begin{array}{rcl}
\ux\cdot\~E_s &=&\ux\cdot\~E_{e}\\
\uy\cdot\~E_s &=&\uy\cdot\~E_{e}\\
\ux\cdot\~H_s &=&\ux\cdot\~H_{e}\\
\uy\cdot\~H_s &=&\uy\cdot\~H_{e} \\
 \end{array}\label{eqn:BCs}
 \right\}\,.
\end{equation}
These equations may be cast in matrix form as
\begin{equation}
\bar{M}\cdot \left( \begin{array}{l} A_{s1}\\ A_{s2}\\ A_{e1}
\\ A_{e2}\end{array}\right) = \left(
\begin{array}{l} 0\\ 0\\ 0 \\ 0\end{array}\right)\,,
\end{equation}
where $\bar{M}$ is a 4$\times$4 matrix.  For a non-trivial solution, the
determinant of  $\bar{M}$ must equal zero; thus, the SWP dispersion equation is
\begin{equation}
{\rm det} \,\bar{M} = 0\,. \label{SWPd}
\end{equation}
Because of the complexity of Equation~(\ref{SWPd}), an algebraic result could not be obtained and recourse
was taken to a numerical method of solution. Parenthetically, when $\vert\#E^{dc}\vert=0$, the Pockels effect is not invoked, and the presented formulation simplifies to that of
Polo et al. (2007a).

\section{Numerical Results and Discussion}
The Pockels effect occurs only in dielectric materials that lack inversion symmetry. Some examples of materials of
this type are potassium niobate, lithium niobate and gallium arsenide. We chose potassium niobate
for  calculations since it has very large EO coefficients. As potassium niobate
belongs to the orthorhombic $mm2$ class, the only non-zero EO
coefficients are $r_{13}$, $r_{23}$, $r_{33}$, $r_{42}$, and $r_{51}$; hence,
we get
\begin{equation}
\displaystyle{\bar{\epsilon}_{PE}\approx\left(
\begin{array}{ccc}
\epsilon _{1}^{(0)}(1-\epsilon _{1}^{(0)}\, r_{13}E_{3}^{dc} ) &
0 &
-\epsilon _{1}^{(0)}\epsilon _{3}^{(0)}\, r_{51}E_{1}^{dc} \\[5pt]
0 &
\epsilon _{2}^{(0)}(1-\epsilon _{2}^{(0)}\, r_{23}E_{3}^{dc} ) &
-\epsilon _{2}^{(0)}\epsilon _{3}^{(0)}\, r_{42}E_{2}^{dc} \\[5pt]
-\epsilon _{3}^{(0)}\epsilon _{1}^{(0)}\, r_{51}E_{1}^{dc} &
-\epsilon _{3}^{(0)}\epsilon _{2}^{(0)}\, r_{42}E_{2}^{dc} &
\epsilon _{3}^{(0)}(1-\epsilon _{3}^{(0)}\, r_{33}E_{3}^{dc} )
\end{array}
\right) }\,  \label{PocEps-mm2}
\end{equation}
from Equation (\ref{PocEps}). Constitutive data for potassium niobate  are as follows (Zgonik et al.,
1993): $\epsilon_1^{(0)} = 4.72$, $\epsilon_2^{(0)}= 5.20$, $\epsilon_3^{(0)}=5.43$, $r_{13}=34\times
10^{-12}$~m~V$^{-1}$, $r_{23}=6\times 10^{-12}$~m~V$^{-1}$, $r_{33}=63.4\times 10^{-12}$~m~V$^{-1}$,
$r_{42}=450\times 10^{-12}$~m~V$^{-1}$, and $r_{51}=120\times 10^{-12}$~m~V$^{-1}$.

The existence of a surface wave was determined at various values of $\chi$, $\psi$, and $n_s$  by satisfactory solution of Equation (\ref{eqn:BCs}).  Purely for illustrative purposes, the most detailed calculations were performed at only one value of   $\chi$: $60^\circ$; at this value of $\chi$ the range of propagation directions is relatively wide.  Let us note that the range of values of $n_s$ for SWP at the planar interface of a non-EO biaxial dielectric material and a non-EO isotropic material is limited (Polo et al., 2007), and the range of values of  the orientation angle  $\psi$ for a specific $n_s$ is quite small as well. In the remainder of this section, the mid-point of the $\psi$-range is denoted by $\psi_m$ and the width of that range by $\Delta\psi$.

\subsection{Results for $\chi=60^\circ$}
\begin{figure}
    \begin{center}
    \begin{tabular}{c}
    \includegraphics[width=7cm]{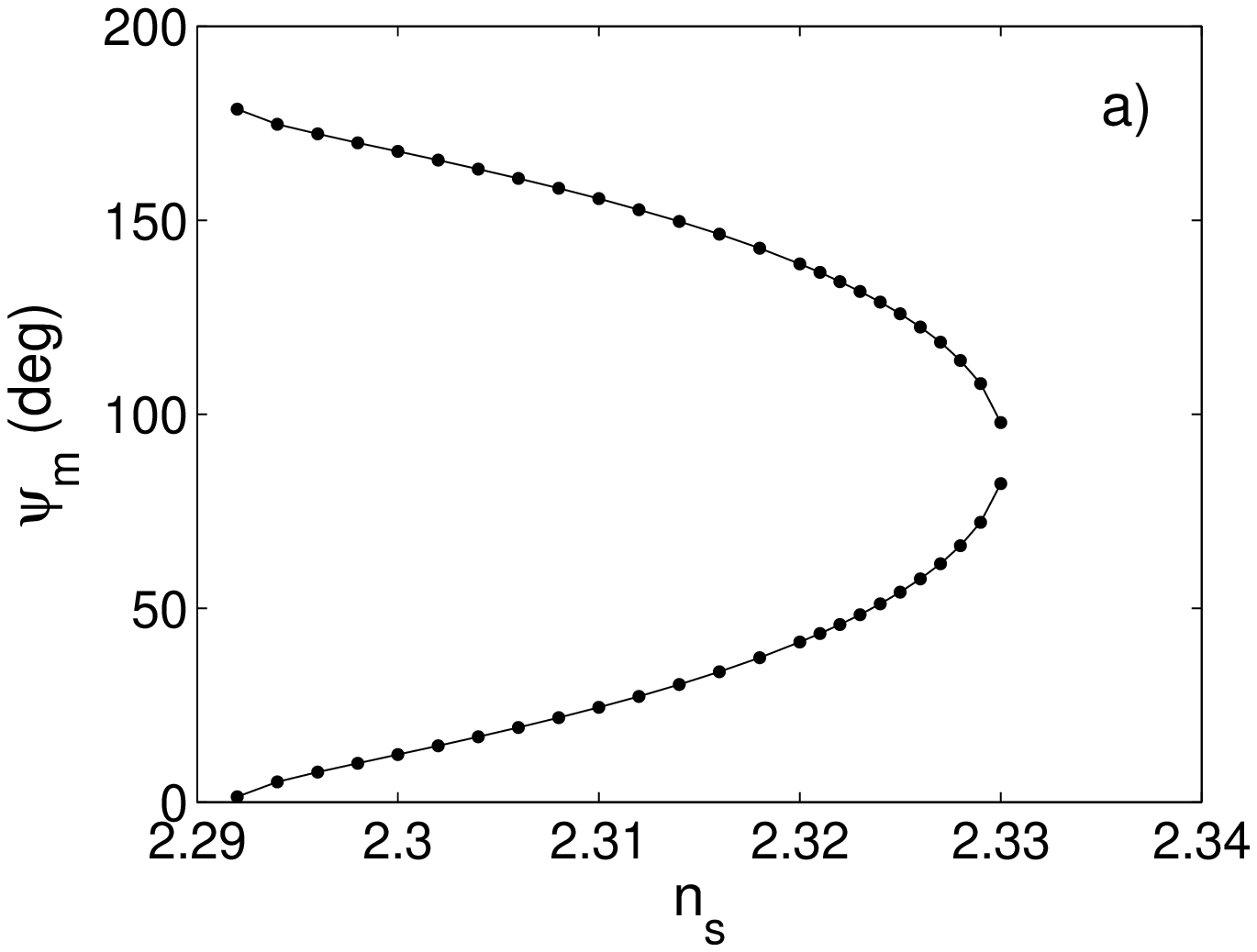}
    \includegraphics[width=7cm]{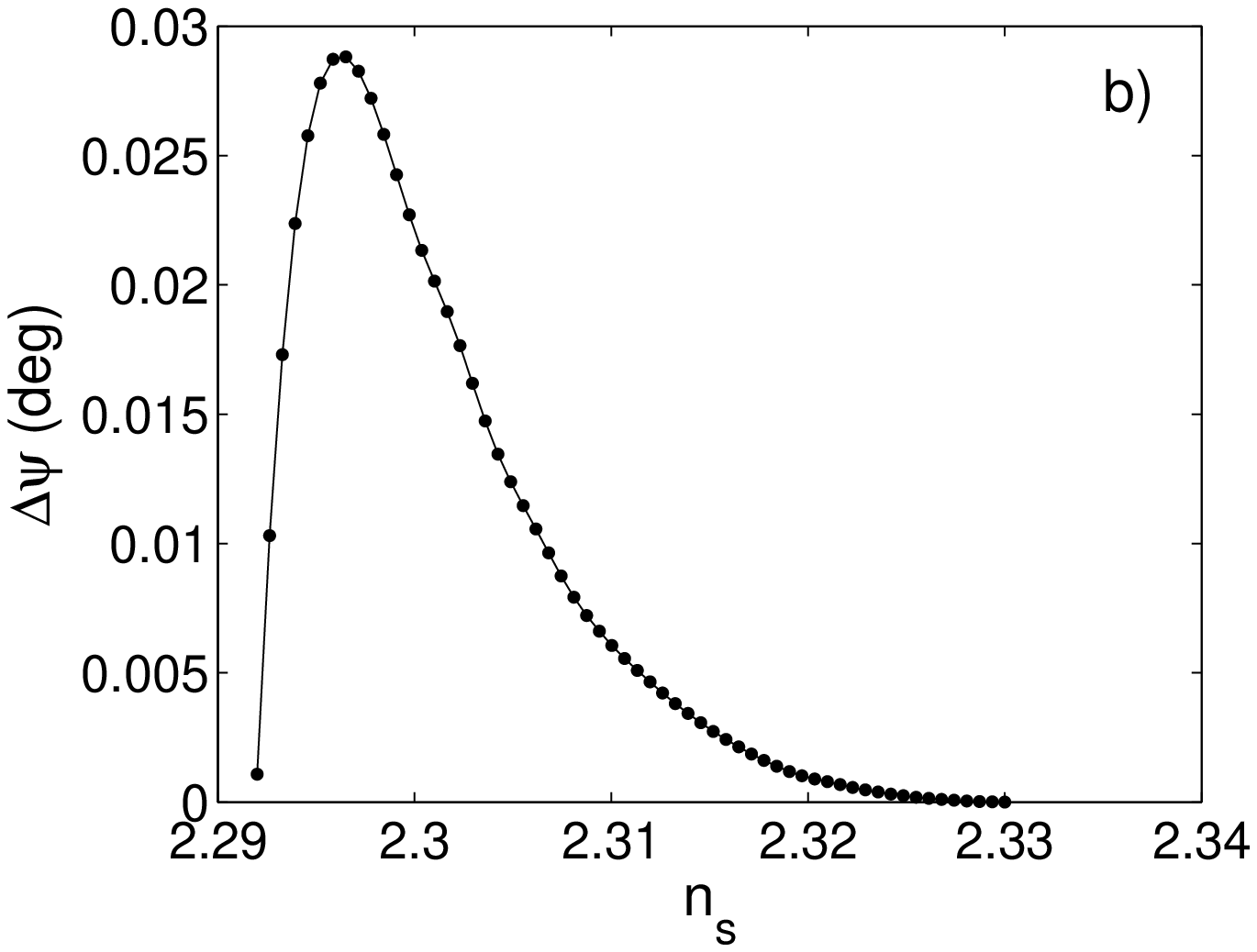}
    \end{tabular}
    \end{center}
    \caption{ \label{Fig:EdcZero} Median propagation angle $\psi_m$ and the width  $\Delta\psi$ of the $\psi$-range, when $\chi=60^\circ$ and $\vert\#E^{dc}\vert=0$. See the text for other parameters.}
 \end{figure}

In order to provide a baseline for the influence of the Pockels effect on SWP, Figure~\ref{Fig:EdcZero}
shows $\psi_m$ and $\Delta\psi$ as  functions of $n_s$ for $\chi=60^\circ$ when $\vert\#E^{dc}\vert=0$.
The plots in the figure cover the entire  $n_s$-range over which SWP was found possible.  The
$n_s$-range extends from approximately $2.292$ to $2.33$, and is thus only $0.038$ in width. The
$\psi$-range in the figure is limited to $[0^\circ,180^\circ]$;  if
SWP is possible for a certain value of $\psi$, it is also possible for $- \psi$, when
$\vert\#E^{dc}\vert=0$.

Over the range $0^\circ\leq \psi\leq 180^\circ$ two bands of $\psi$ values for SWP
can be deduced from
 Figure~\ref{Fig:EdcZero} for each value of $n_s$, except at the largest value
of $n_s$ where the two bands coalesce. At the lower limit of the $n_s$-range, $\psi_m$ approaches either
$0^\circ$ or $180^\circ$; while at the upper limit of that range, $\psi_m$ approaches $90^\circ$ for
both bands. When one considers the entire  $\psi$-range (i.e., $-180^\circ\leq \psi\leq 180^\circ$),
there are four bands of $\psi$-values that merge into two bands at both limits of the
$n_s$-range.

In Figure~\ref{Fig:EdcZero}b, $\Delta\psi$ is shown as a function of $n_s$.  Only one curve is shown
since both $\psi$-bands have the same width at each value of $n_s$.  The curve has a single peak and $\Delta\psi$
goes to zero at the two endpoints of the $n_s$-range. The maximum value of $\Delta\psi$ is about
$0.03^\circ$ and occurs at $n_s\approx2.296$, i.e., near the lower endpoint of the $n_s$-range.
Thus,   the   $\psi$--bands for SWP are really narrow.

In order to explore the influence of the Pockels effect on SWP, $\vert\#E^{dc}\vert$
was next set equal to
$10^7$~V~m$^{-1}$.  This dc field was oriented along each of the laboratory coordinate
axes $(x,y,$ and $z$) separately.  We now describe the results for each direction of orientation of
the dc electric field in order of increasing complexity.

\begin{figure}
    \begin{center}
    \begin{tabular}{c}
    \includegraphics[width=7cm]{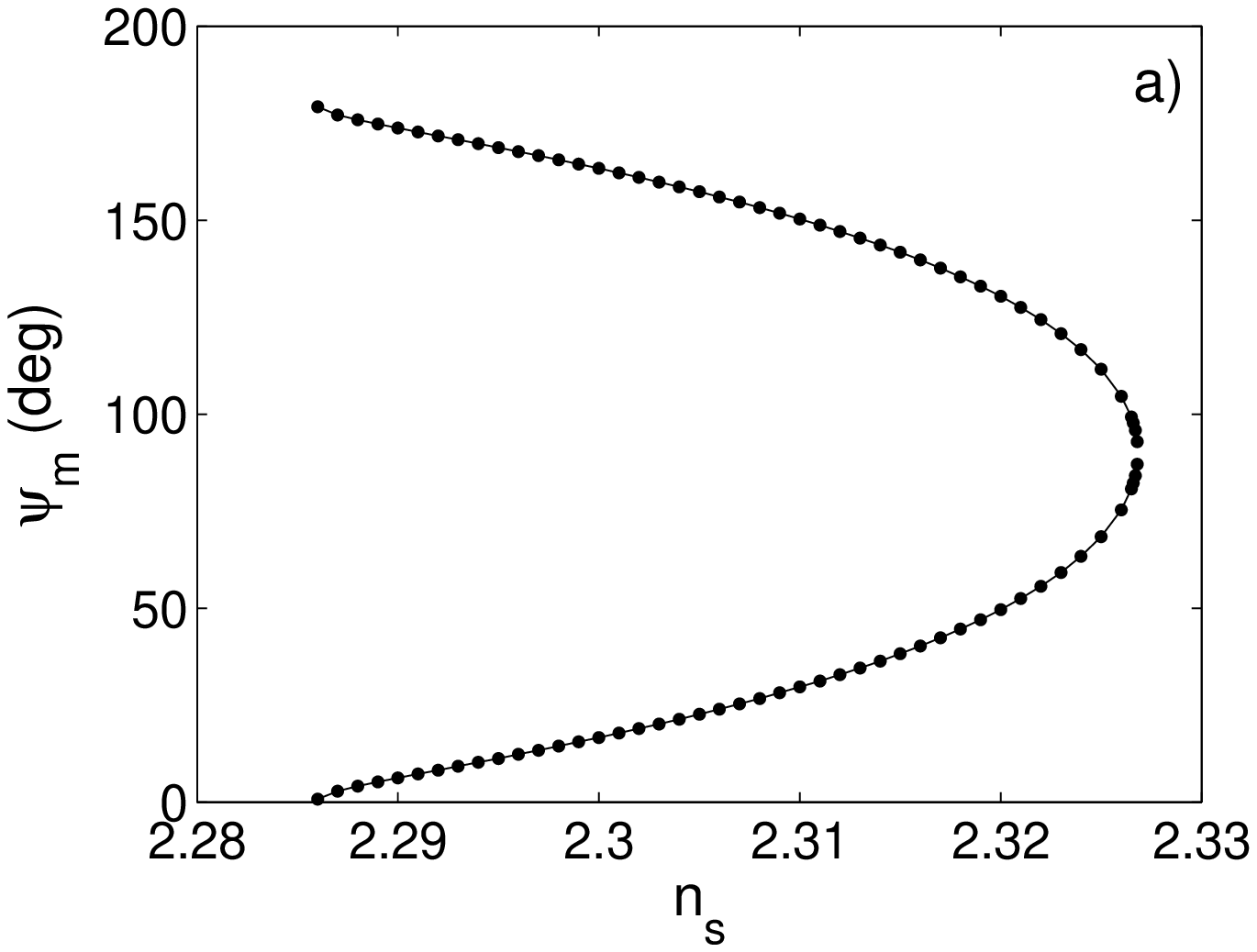}
    \includegraphics[width=7cm]{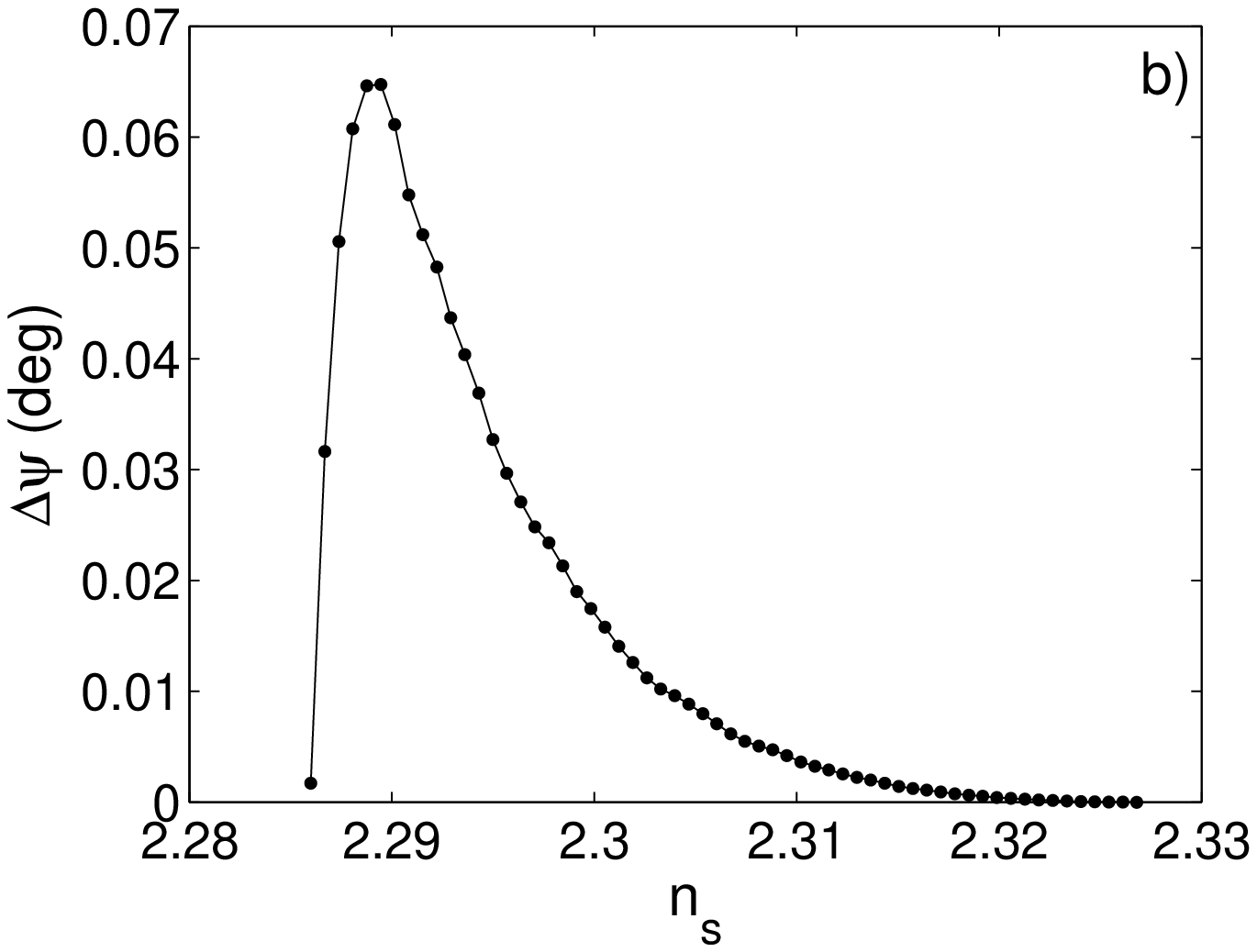}
    \end{tabular}
    \end{center}
    \caption{Same as Figure~\ref{Fig:EdcZero}, except that $\#E^{dc}=\uz\,10^7$~V~m$^{-1}$.
    }\label{Fig:Ez}
 \end{figure}

Let us begin with $\#E^{dc}=\uz\,10^7$~V~m$^{-1}$. Both $\psi_m$ and $\Delta\psi$ are plotted against
$n_s$  in Figure~\ref{Fig:Ez}.  Just as for the plots for $\vert\#E^{dc}\vert=0$ already discussed,
there are four  $\psi$-bands with mirror symmetry about the $n_s$-axis.  Figure~\ref{Fig:Ez}a shows that
the  $n_s$-range for SWP has grown slightly and shifted to lower values of $n_s$ compared to
Figure~\ref{Fig:EdcZero}a.  With approximate lower and upper endpoints of 2.286 and 2.327 respectively, the
width of the $n_s$-range is 0.041. Both $\psi$-bands meet at $\psi=90^\circ$, which is just the same as
in the absence of the dc electric field. In Figure \ref{Fig:Ez}b, a single curve describes the widths of
both $\psi$-bands as $n_s$ varies, which is similar to the curve in  Figure~\ref{Fig:EdcZero}b for
$\vert\#E^{dc}\vert=0$. The height of the $\Delta\psi$-peak, however, is approximately $0.062^\circ$, a
little more than double that of the peak for $\vert\#E^{dc}\vert=0$, and occurs at $n_s\approx2.289$.

\begin{figure}
    \begin{center}
    \begin{tabular}{c}
    \includegraphics[width=7cm]{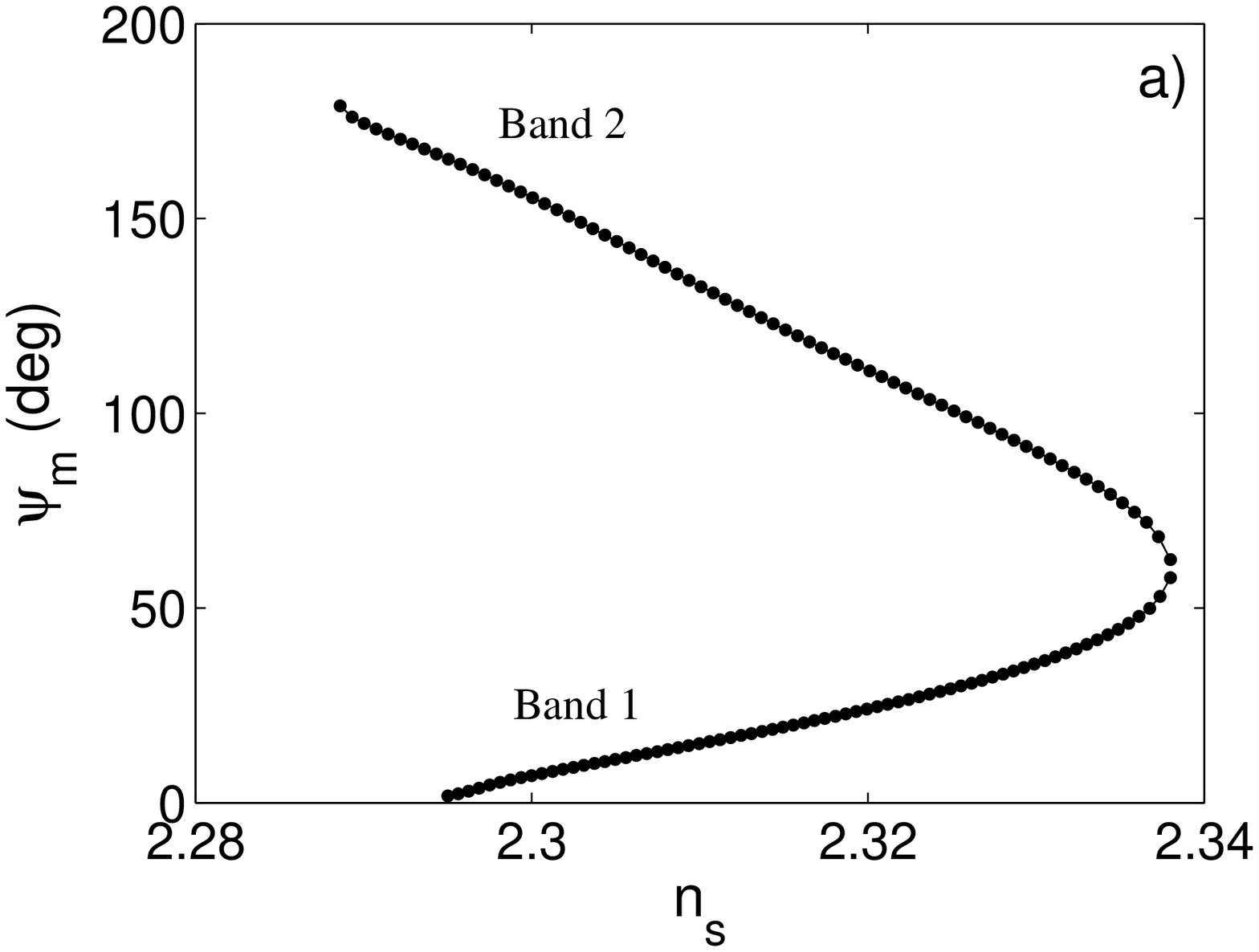}
    \includegraphics[width=7cm]{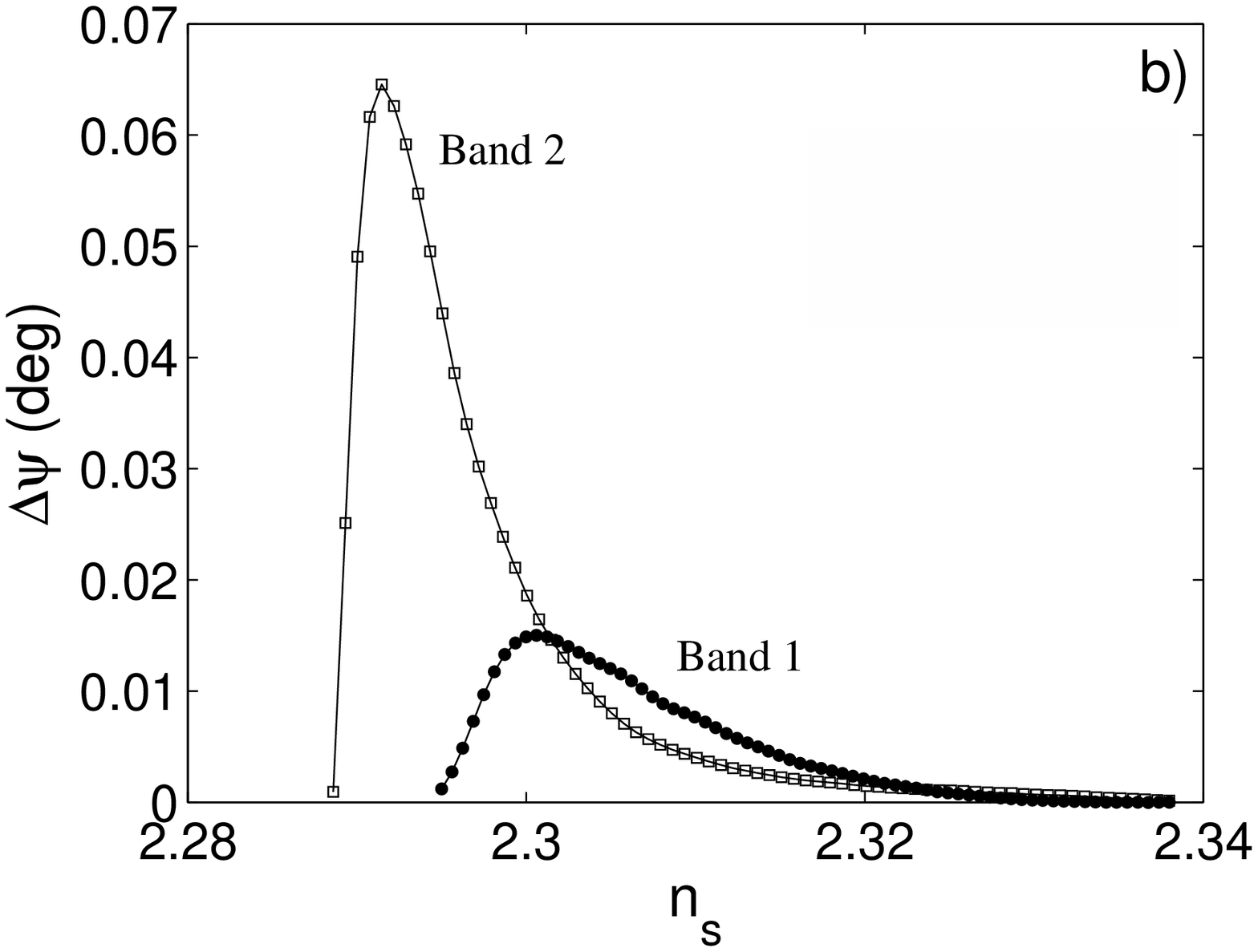}
    \end{tabular}
    \end{center}
    \caption{ Same as Figure~\ref{Fig:EdcZero}, except that $\#E^{dc}=\ux\,10^7$~V~m$^{-1}$.}\label{Fig:Ex}
 \end{figure}

Figure \ref{Fig:Ex} shows the $\psi_m$-$n_s$ and $\Delta\psi$-$n_s$ curves when
$\#E^{dc}=\ux\,10^7$~V~m$^{-1}$. Just as in the previous two cases, there are four $\psi$-bands  with
mirror symmetry about the $n_s$-axis. So, only two $\psi$-bands are shown in the figure. Evidently from
Figure~\ref{Fig:Ex}a, the $n_s$-range for SWP is larger than when the dc electric field is either absent
or aligned parallel to the $z$ axis. The width of the  $n_s$-range for SWP has grown to 0.049,
with the lower and upper endpoints of the $n_s$-range being 2.289 and 2.338, respectively. Although both
$\psi$-bands in Figure \ref{Fig:Ex}a have wider $n_s$-ranges than in the previous two figures,
 the upper band (labeled Band 2), has increased more than the lower band (Band 1). In addition,
the two $\psi$-bands meet at a lower value of $\psi_m$  ($\approx60^\circ$), and, thus, lack the
symmetry about $\psi_m=90^\circ$ seen in Figures \ref{Fig:EdcZero}a and \ref{Fig:Ez}a for the cases of
$\#E^{dc}=0$ and $\#E^{dc}\parallel\uz$, respectively.

Both $\psi$-bands in Figure \ref{Fig:Ex}b still show a single $\Delta\psi$-peak each, towards the lower
endpoint of the $n_s$-range. However, the $\Delta\psi$-$n_s$ curves for the two bands are not identical.
Whereas the $\Delta\psi$-peak for Band 2 is about the same as for $\#E^{dc}\parallel\uz$ in Figure
\ref{Fig:Ez}b, the $\Delta\psi$-peak for Band 1 is about a third of that value. In addition, the
positions of the $\Delta\psi$-peaks are shifted slightly from the value for $\vert\#E^{dc}\vert=0$, with
the peak for Band 2 shifted downward to $n_s=2.300$ and that for Band 1 upward to $n_s=2.292$.

\begin{figure}
    \begin{center}
    \begin{tabular}{c}
    \includegraphics[width=7cm]{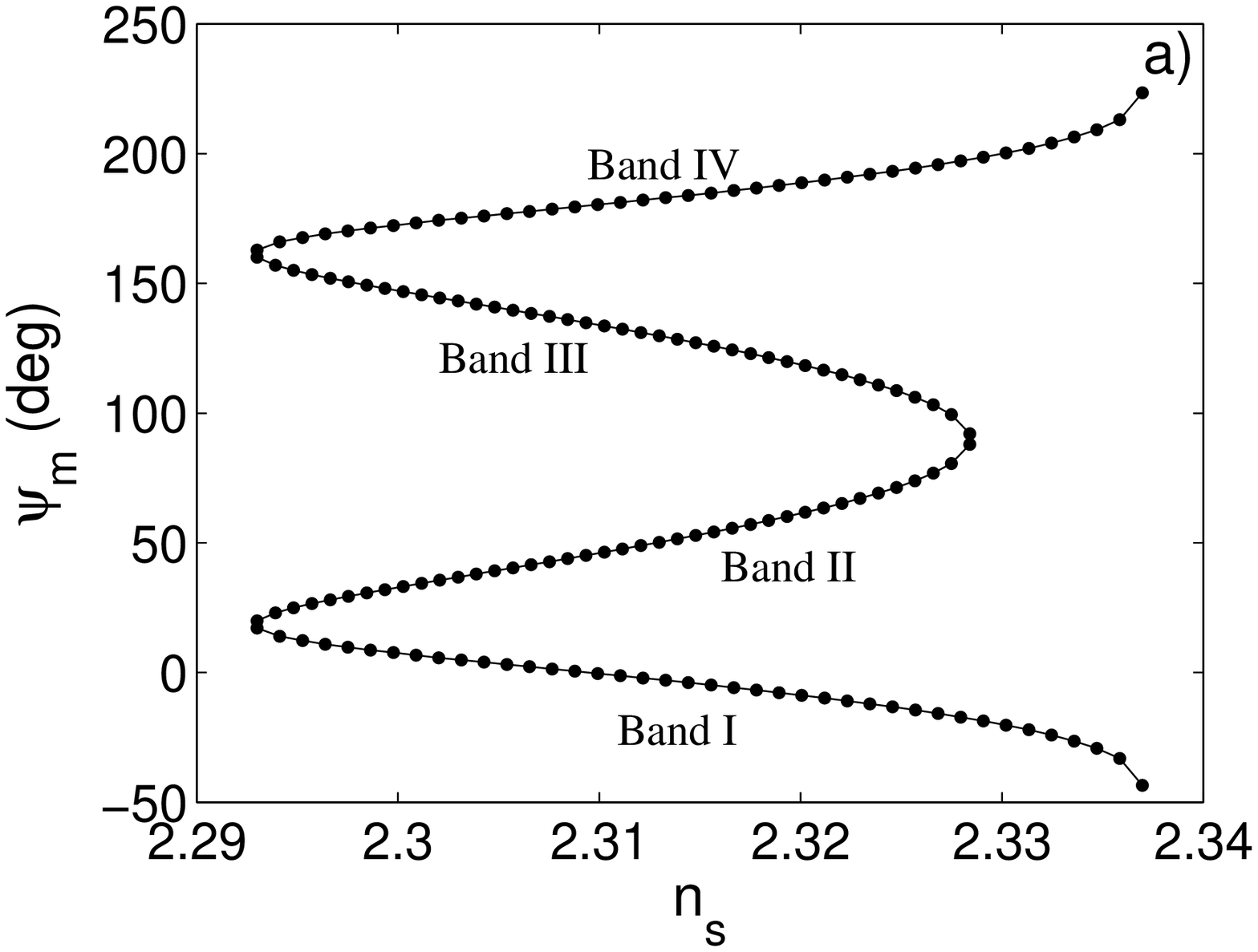}
    \includegraphics[width=7cm]{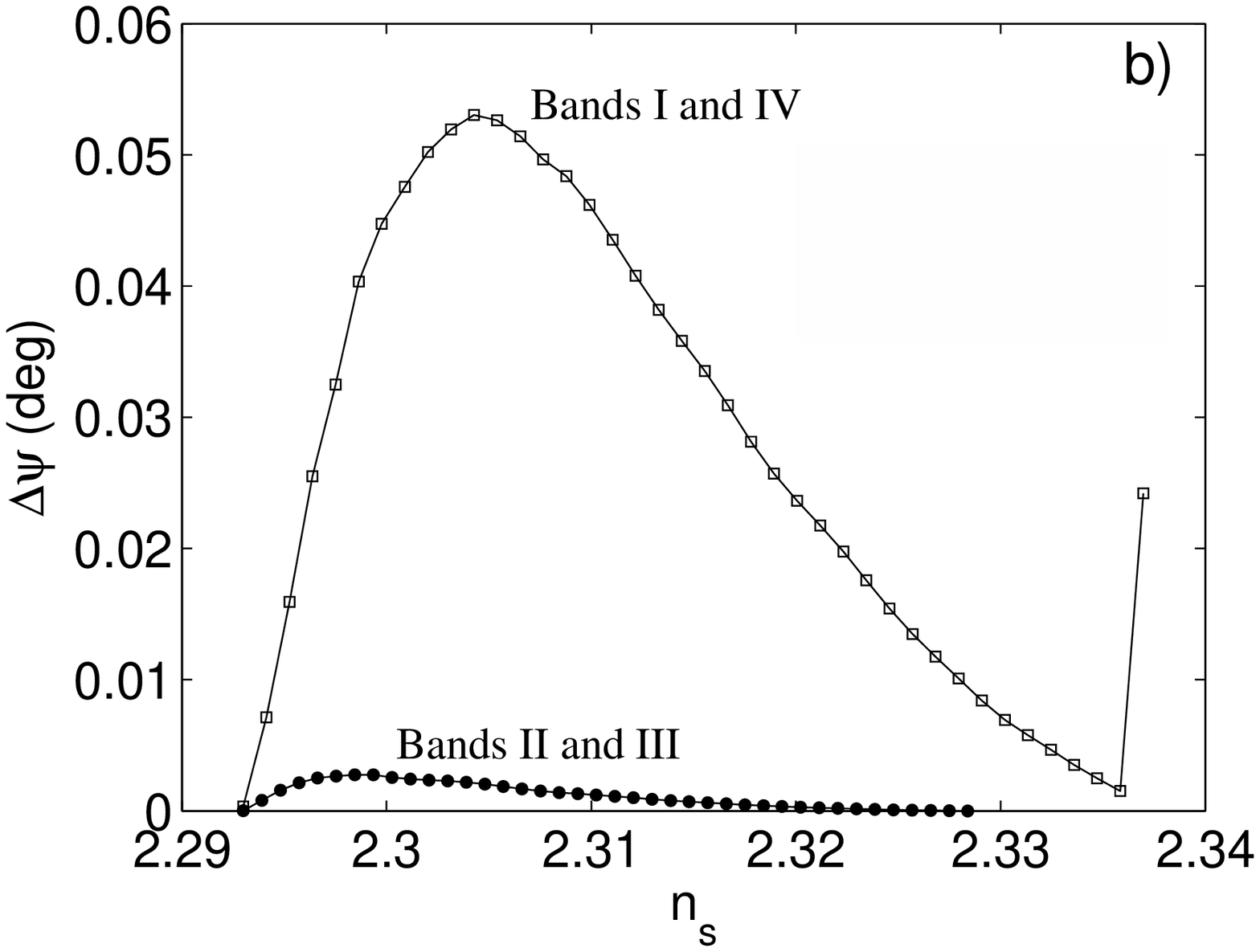}
    \end{tabular}
    \end{center}
    \caption{ Same as Figure~\ref{Fig:EdcZero}, except that $\#E^{dc}=\uy\,10^7$~V~m$^{-1}$. }\label{Fig:Ey}
 \end{figure}

Figure \ref{Fig:Ey} displays the influence of the dc electric field when applied along the $y$ axis,
i.e., $\#E^{dc}=\uy 10^7$~V~m$^{-1}$. In
Figure~\ref{Fig:Ey}a, the full range of $\psi$, $[-180^\circ,180^\circ]$, is displayed as the
$\psi$-bands for SWP are no longer symmetric about $\psi=0^\circ$; instead, these bands (labeled I to IV) are
now located symmetrically about $\psi=90^\circ$.  The widths $\Delta\psi$ of the $\psi$-bands are shown
in Figure \ref{Fig:Ey}b.  As is the case for the dc electric field applied parallel to the $x$ axis, the
widths of all four bands are not equal. Band I with a $\psi$-range of approximately
$[-27^\circ,17^\circ]$ and Band IV with an approximate $\psi$-range of $[163^\circ,233^\circ]$ have the
same width as a function of $n_s$. The maximum width of these two $\psi$-bands is about $0.052^\circ$
and occurs near $n_s=2.304$. Similarly, Band II with a $\psi$-range of $[20^\circ,84^\circ]$ and Band
III with a $\psi$-range of $[96^\circ,160^\circ]$ share a common $\Delta\psi$-$n_s$ curve. The widths of
Bands II and III  are more than a factor of 10 smaller than of Bands I and IV, the $\Delta\psi$-peak for
Bands II and III being $0.0028^\circ$ at $n_s\approx2.299$.

The sudden change in $\Delta\psi$ at the upper endpoint of the $n_s$ range for Bands I and IV in
Figure~\ref{Fig:Ey}b should be noted. A similar sudden change was also found for $\chi=75^\circ$. The
region of coalescence of two bands on the $n_s$-axis is hard to delineate accurately. Quite possibly,
the sudden change is a numerical artifact; it will be the subject of future investigations.

\begin{figure}
    \begin{center}
    \begin{tabular}{c}
    \includegraphics[height=7cm]{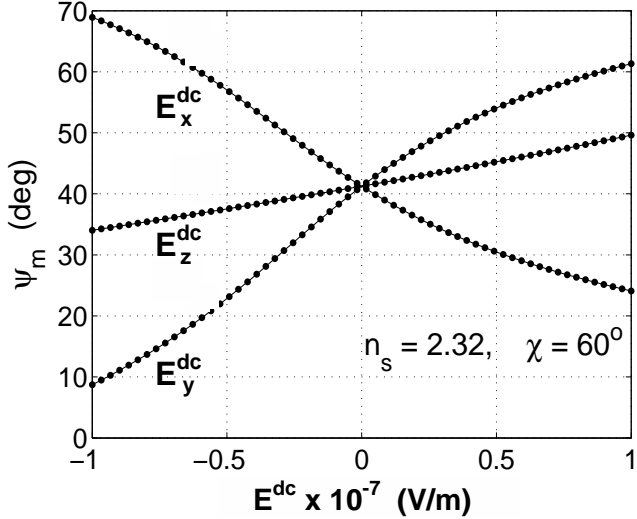}
    \end{tabular}
    \end{center}
    \caption{ Median propagation angle $\psi_m$ vs. the signed magnitude
    of $\#E^{dc}$ for $\#E^{dc}$ parallel to the $x$, $y$, and $z$ axes;
    $n_s=2.32$, $\chi=60^\circ$. See the text for other parameters.
    }\label{Fig:Field_Variation}
 \end{figure}

The effect of variation of the magnitude of the dc electric field is shown in Figure
\ref{Fig:Field_Variation} with a plot of $\psi_m$ vs. the dc field's signed magnitude, for $\#E^{dc}$
parallel to the $x$, $y$, and $z$ axes. For the dc field oriented along both the $y$  and the $z$ axes,
$\psi_m$ increases as the signed magnitude becomes more positive:   the relationship is nearly linear
for $\#E^{dc}\parallel\uz$, and somewhat S-shaped for $\#E^{dc}\parallel\uy$.  On the other hand, when
$\#E^{dc}\parallel\ux$, $\psi_m$ decreases as the signed magnitude becomes more positive. The foregoing
results clearly show that by varying the magnitude and/or direction of the dc electric field, the
direction of SWP, relative to the crystal axes of the EO material, can be controlled.

\subsection{Results for other $\chi$}

\begin{figure}
    \begin{center}
    \includegraphics[width=14cm]{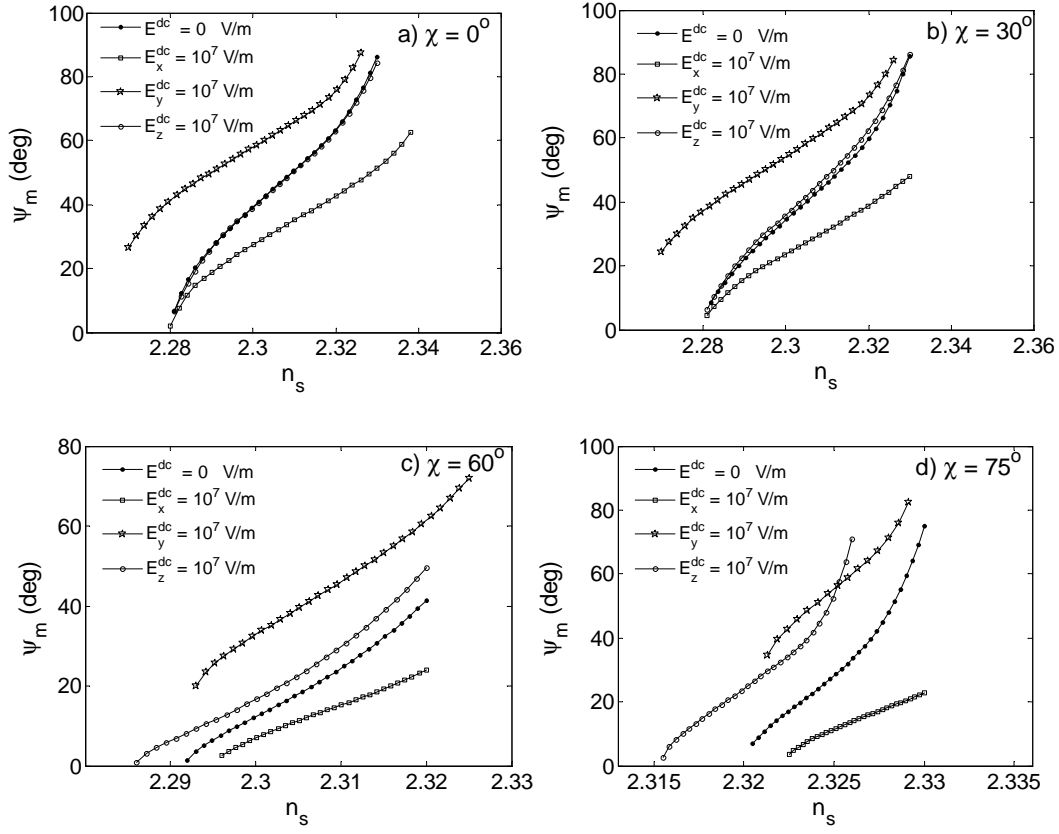}
    \end{center}
    \caption{ Median propagation angle $\psi_m$ versus $n_s$ for: a) $\chi=0^\circ$,\; b) $\chi=30^\circ$,\;
     c) $\chi=60^\circ$,\; d) $\chi=75^\circ$. Only Band 1 is shown for $\#E^{dc}\parallel \ux$;
     only Band II is shown for $\#E^{dc}\parallel \uy$.
     See the text for other parameters.}
    \label{Fig:PsiAllChi}
 \end{figure}

Limited results were also obtained for $\chi=0^\circ$, $30^\circ$, and $75^\circ$, with the search for
SWP restricted to $\psi\in\left[0^\circ,90^\circ\right]$. Figure \ref{Fig:PsiAllChi} contains
$\psi_m$-$n_s$ curves for  $\chi\in\left\{0^\circ, 30^\circ,60^\circ,75^\circ\right\}$, for (a)
$\#E^{dc}=0$ and (b) $\#E^{dc}$ of signed magnitude $+10^7$~V~m$^{-1}$ and parallel to the $x$, $y$, and
$z$ axes. Although both Band 1 and Band 2 (when $\#E^{dc}\parallel \ux$) and Band I and Band II (when
$\#E^{dc}\parallel \uy$) should be visible in the plots, for simplicity, only Band 1 and Band II are
shown.

 The curves for the chosen values of $\chi$ are isomorphic. At each value of
$\chi$, a marked downward shift of the $\psi_m$-$n_s$ curve, compared to the  case for
$\vert\#E^{dc}\vert=0$, occurs when $\#E^{dc}\parallel\ux$, and an upward shift for
$\#E^{dc}\parallel\uy$. There is also a shift for $\#E^{dc}\parallel\uz$, but it is much smaller and is
almost unnoticeable at the lower values of $\chi$; however, the shift is significant for
$\chi=75^\circ$,  on the same order as when the dc electric field is applied along the other two
axes.

\begin{figure}
    \begin{center}
    \includegraphics[width=14cm]{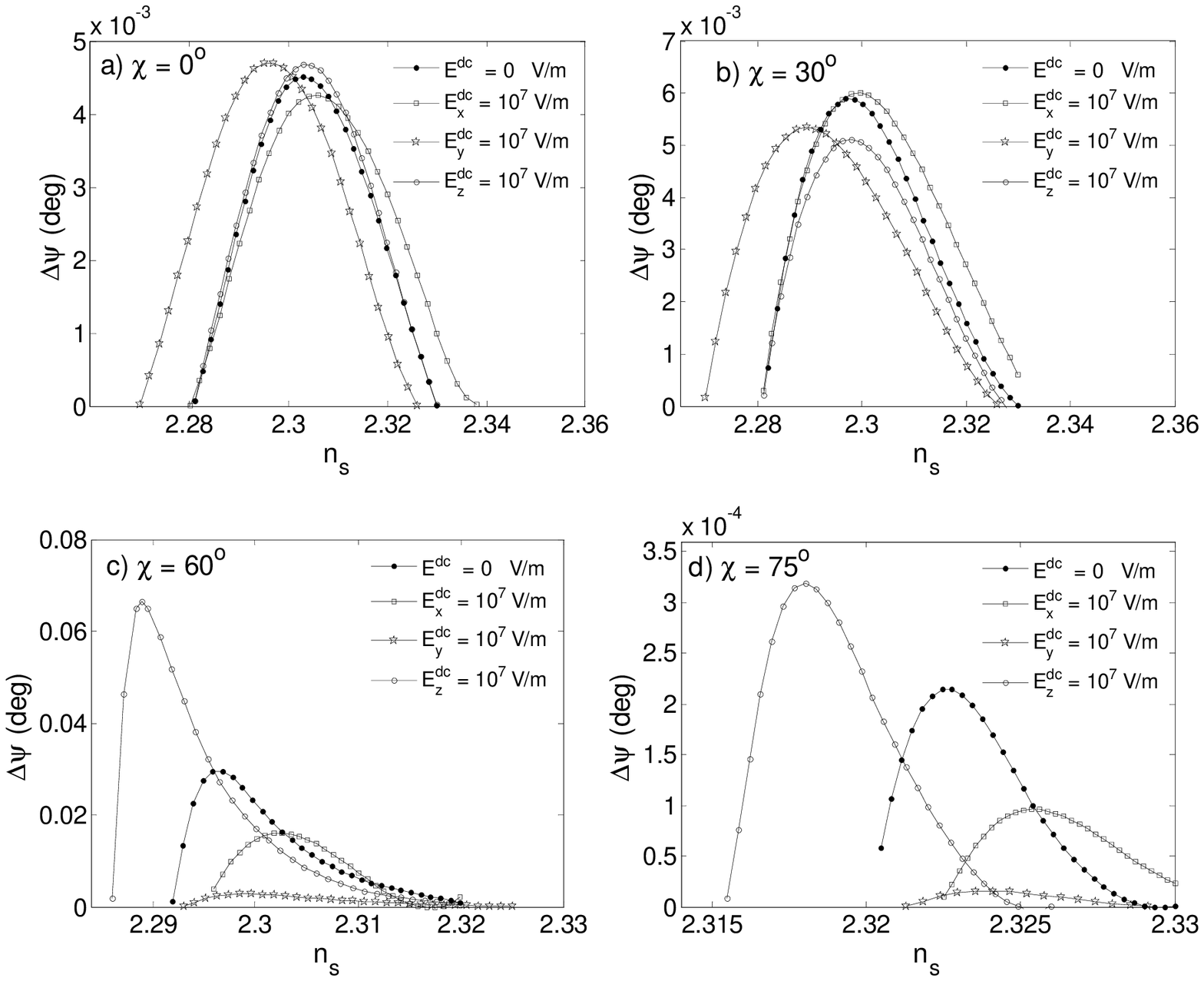}
    \end{center}
    \caption{ Same as Figure~\ref{Fig:PsiAllChi}, except that $\Delta\psi$ is plotted
    against $n_s$.
    }\label{Fig:DeltaPsiAllChi}
 \end{figure}

Figure~\ref{Fig:DeltaPsiAllChi} shows $\Delta\psi$-$n_s$ curves for $\chi\in\left\{0^\circ,
30^\circ,60^\circ,75^\circ\right\}$, for the dc electric field configured as for
Figure~\ref{Fig:PsiAllChi}. When $\chi=0^\circ$, the curves for all three orientations of the dc field
and the zero dc field are similar, having nearly the same peak values of $\Delta\psi$ and peaking at
close to the same values of $n_s$. As $\chi$ increases, the curves become differentiated both in the
$\Delta\psi$ peak height and its position on the $n_s$ axis.  Among the four values of $\chi$ explored,
the largest peak values of $\Delta\psi$ occur for $\chi=60^\circ$. This is particularly true when
$\#E^{dc}\parallel\uz$; then some  values of $\Delta\psi$ are more than an order of magnitude larger
than found at other values of $\chi$.

\section{Concluding remarks}

The influence of the Pockels effect on SWP at the interface between potassium niobate and an isotropic
dielectric material has been demonstrated by our numerical studies.  With the application of a dc
electric field, we have noted the shift in the $n_s$-range and the $\psi$-bands that permit SWP.  The
greatest influence of the Pockels effect is on the median propagation angle  $\psi_m$ of the very narrow
$\psi$-bands in which SWP is possible.  Shifts of over $30^\circ$ at a fixed value of $n_s$ have been
deduced. Thus, the Pockels effect can be pressed into service for an electrically controlled on-off
switch for surface waves.

As of now, the existence of surface waves at the planar interface between a biaxial dielectric material
and an isotropic dieletric material has not been demonstrated experimentally.  The narrow $\psi$- and
$n_s$-regimes for SWP may discourage experimentalists from searching for surface waves in these
scenarios. Electrical control of SWP direction may provide a convenient way to search for surface waves.
The experiment could be carried out with a fixed geometry defining the propagation direction and the
orientation of the linear  EO material, and  the signed magnitude of $\#E^{dc}$ could then easily be
swept electronically until a surface wave is detected. Various experimental configurations for exciting
surface waves are available (Agranovich and Mills, 1982), as also are optically transparent electrodes to apply dc
electric fields (Minami 2005; Medvedeva 2007). Finally, we hope that our work shall spur the development of
artificial linear EO materials with much higher EO coefficients than now available.

\bigskip
{\bf References}
\begin{itemize}
\item[]
Agranovich, V. M., \& D. L. Mills. (Eds.) 1982. \emph{Surface Polaritons:
Electromagnetic Waves at Surfaces and
Interfaces}. Amsterdam:
North-Holland.

\item[]
Averkiev, N. S., \& M. I. Dyakonov. 1990. Electromagnetic waves
localized at the interface of transparent anisotropic media.
\emph{Opt. Spectrosc. (USSR)}  68:653-655.

\item[]
Boardman, A. D. (Ed.) 1982. \emph{Electromagnetic Surface Modes}. Chichester: Wiley.

\item[]
Boyd, R. W. 1992. \emph{Nonlinear Optics}. San Diego:
Academic.

\item[]
Chen, H. C. 1983. \emph{Theory of Electromagnetic Waves: A
Coordinate-Free Approach}, 219-226. New York: McGraw-Hill.

\item[]
Cook, Jr., W. C. 1996.  Electrooptic coefficients. In: Nelson, D.F.   (Ed.) 1996.
\emph{Landolt-Bornstein, Vol. 3/30A}, 164.  Berlin: Springer.

\item[]
Darinski\u{i}, A. N. 2001.
Dispersionless polaritons on a twist boundary in optically uniaxial
crystals. \emph{Crystallogr. Repts.} 46:842-844.

\item[]
D'yakonov, M. I. 1988. New type of electromagnetic wave propagating at
an interface. \emph{Sov. Phys. JETP} 67:714-716.

\item[]
Farias, G. A., E. F. Nobre, \& R. Moretzsohn. 2002. Polaritons in
hollow cylinders in the presence of a dc magnetic field. \emph{J.
Opt. Soc. Am. A} 19:2449-2455.

\item[]
Homola, J.,  S. S. Yee, \& G. Gauglitz. 1999. Surface plasmon resonance
sensors:  review. \emph{Sens. Actuat. B: Chem.} 54:3-15.

\item[]
Lakhtakia, A. 2006a. Electrically switchable exhibition of circular
Bragg phenomenon by an isotropic slab. \emph{Microw. Opt. Technol.
Lett.} 48:2148-2153; corrections: 2007. 49:250-251.

\item[]
Lakhtakia, A. 2006b. Narrowband and ultranarrowband filters with electro-optic
structurally chiral materials. \emph{Asian J. Phys.} 15:275-282.

\item[]
Lakhtakia, A., \& T. G. Mackay. 2007.
Electrical control of the linear optical
properties of particulate composite materials.
\emph{Proc. R. Soc. Lond. A} 463:583-592.

\item[]
Lakhtakia, A., \& J. A. Reyes. 2006a.
Theory of electrically controlled exhibition of circular Bragg phenomenon by an obliquely excited structurally chiral material~--~Part 1: Axial dc electric field.
\emph{Optik} doi:10.1016/j.ijleo.2006.12.001.

\item[]
Lakhtakia, A., \& J. A. Reyes. 2006b.
Theory of electrically controlled exhibition of circular Bragg phenomenon by an obliquely excited structurally chiral material~--~Part 2: Arbitrary dc electric field.
\emph{Optik} doi:10.1016/j.ijleo.2006.12.002.

\item[]
 Li, J., M.-H. Lu, L. Feng, X.-P. Liu, \& Y.-F. Chen. 2007.
 Tunable negative refraction based on the Pockels effect in two-dimensional
photonic crystals composed of electro-optic crystals. \emph{J. Appl. Phys.}
 101:013516.

\item[]
Mackay, T. G., \& A. Lakhtakia. 2007.
Scattering loss in electro-optic particulate composite materials.
\emph{J. Appl. Phys.} 101:083523.

\item[]
Matveeva, E., Z. Gryczynski, J. Malicka, J. Lukomska, S. Makowiec,
K. Berndt, J. Lakowicz, \& I. Gryczynski. 2005. Directional surface
plasmon-coupled emission: Application for an immunoassay in whole
blood. \emph{Anal. Biochem.} 344:161-167.

\item[]
Medvedeva, J. E. 2007.
Unconventional approaches to combine
optical transparency with electrical conductivity. \emph{Appl. Phys. A}
doi:10.1007/s00339-007-4035-4.

\item[]
Minami, T. 2005.
Transparent conducting oxide semiconductors for transparent electrodes.
\emph{Semicond. Sci. Technol.} 20:S35-S44.

\item[]
Mineralogy Database,
http://www.webmineral.com/ (20 April 2006).

\item[]
Nelatury, S. R., J. A. Polo Jr., \& A. Lakhtakia.
2007.
Surface waves with simple exponential transverse decay at a biaxial bicrystalline interface.
\emph{J. Opt. Soc. Am. A} 24:856-865; corrections: 24:2102.

\item[]
Polo Jr., J. A., S. Nelatury, \& A. Lakhtakia. 2006. Surface electromagnetic wave at a tilted uniaxial
bicrystalline interface. \emph{Electromagnetics} 26:629-642.

\item[]
Polo Jr., J. A., S. R. Nelatury, \& A. Lakhtakia. 2007a. Propagation of
surface waves at the planar interface of a columnar thin film and an
isotropic substrate. \emph{J. Nanophoton.} 1, 013501.

\item[]
Polo Jr., J. A., S. R. Nelatury, \& A. Lakhtakia.
2007b.
Surface waves at a biaxial bicrystalline interface.
\emph{J. Opt. Soc. Am. A} 24:xxxx-xxxx (at press).

\item[]
Torner, L., J. P. Torres, F.~Lederer, D.~Mihalache, D.~M.~Baboiu, \&
M.~Ciumac. 1993. Nonlinear hybrid waves guided by birefringent
interfaces. \emph{Electron. Lett.} 29:1186-1188.

\item[]
Torreri, P.,  M. Ceccarini, P. Macioce, \& T. Petrucci. 2005.
Biomolecular interactions by surface plasmon
resonance technology. \emph{Ann. Ist. Super. Sanit\`{a}} 41:437-441.

\item[]
Walker, D. B., E. N. Glytsis, \& T. K. Gaylord. 1998. Surface mode at
isotropic-uniaxial and isotropic-biaxial
interfaces. \emph{J. Opt. Soc. Am. A} 15:248-260.

\item[]
Wong, C., H. Ho, K. Chan, S. Wu, \& C. Lin. 2005. Application of
spectral surface plasmon resonance to gas
pressure sensing. \emph{Opt. Eng.} 44:124403.

\item[]
Zenneck, J. 1907. \"{U}ber die Fortpflanzung ebener
elektromagnetischer Wellen l\"{a}ngs einer ebenen
Lieterfl\"{a}che und ihre Beziehung zur drahtlosen Telegraphie.
\emph{Ann. Phys. Lpz.} 23:846-866.

\item[]
Zgonik, M.,  R. Schlesser, I. Biaggio, E. Volt, J. Tscherry, \& P. G\"unter. 1993. Material constants of
KNbO$_3$ relevant for electro- and acousto-optics. \emph{J. Appl. Phys.}
74:1287-1297.

\end{itemize}

\end{document}